\def\chng{}

\def\msol{{\rm M}_{\odot}}
\def\nurot{\nu_{\rm rot}}
\documentclass{aa}
\usepackage{graphicx}
\begin{document}
%
%
%
\newcommand{\ddp}[2]{\frac{\partial #1}{\partial #2}}
\newcommand{\ddps}[2]{\frac{\partial^2 #1}{\partial #2 ^2}}

\thesaurus{06(02.04.1, 02.05.2, 02.07.1, 08.14.1, 13.25.5)}
\title{Innermost stable circular orbits around strange stars\\
         and kHz QPOs in low-mass X-ray binaries }
\author{J. L. Zdunik \inst{1}
 \and
P. Haensel\inst{1,2}
 \and
D. Gondek-Rosi{\'n}ska\inst{1}
 \and
E. Gourgoulhon\inst{2}
}
\institute{N. Copernicus Astronomical Center, Polish
           Academy of Sciences, Bartycka 18, PL-00-716 Warszawa, Poland
\and
D\'epartement d'Astrophysique Relativiste et de Cosmologie
-- UMR 8629 du CNRS, Observatoire de Paris, F-92195 Meudon Cedex,
France\\
{\em jlz@camk.edu.pl,  haensel@camk.edu.pl, dorota@camk.edu.pl,
  Eric.Gourgoulhon@obspm.fr}}
\offprints{J.L. Zdunik}
\date{received/accepted}
\titlerunning{Innermost stable orbits around strange stars and QPOs}
\authorrunning{J.L.~Zdunik et al.}
\maketitle
%
\begin{abstract}
Exact calculations of innermost stable circular orbit (ISCO)
around rotating strange stars are performed within the framework
of general relativity. Equations of state (EOS) of strange quark
matter based on the MIT Bag Model with massive strange quarks and
lowest order QCD interactions, are used. The presence of a solid crust
of normal matter on rotating, mass accreting strange stars  in
LMXBs  is taken into account. It is found that, contrary to neutron
stars, above some minimum mass (which for the
considered equations of state ranged from $1.4~{\rm M}_\odot$
to $1.6~{\rm M_\odot}$)
  a gap always separates the ISCO and stellar surface,
independently of the strange star rotation rate.
For a given baryon mass of strange star, we calculate the ISCO frequency
as function of stellar rotation frequency, from static to Keplerian
configuration. For masses close to the maximum mass of
static configurations the ISCO frequencies  for static
and  Keplerian configurations are similar. However, for masses
significantly lower than  the maximum mass of static configurations,
the minimum value of the ISCO frequency is reached in the Keplerian limit.
 Presence of a
solid crust increases the ISCO frequency for the Keplerian configuration
by about  ten percent compared to that for a bare strange star of the
same mass. For standard parameters of  strange quark matter EOS,
resulting in a maximum static mass of $1.8~{\rm M}_\odot$,
the ISCO frequency  for $\ga 1.4~{\rm M}_\odot$ strange stars always exceeds
1.07~kHz (the upper QPO frequency  reported for 4U 1820-30).
We give an example of
strange quark matter model, which yields maximum static mass of
$2.3~{\rm M}_\odot$, and for which the ISCO frequency of 1.07 kHz
 is allowed at stellar rotation rates
200-300 Hz, provided the strange star mass exceeds 2.2 ${\rm
M}_\odot$. For this EOS
 even lower  value $\nu_{\rm ISCO}\simeq 1$~kHz is
reached near  the
Keplerian limit,
for a broad range of stellar masses.
  While reproducing $\nu_{\rm
ISCO}=1.07$ kHz at slow rotation rates requires  tuning of
strange quark matter parameters, no such a tuning is required to
reproduce orbital frequencies around strange stars equal to
highest observed upper QPO frequencies.


\end{abstract}
\section{Introduction}
%
Observations of quasi periodic oscillations (QPOs) in the X-ray
fluxes from low-mass X-ray binaries (LMXB), which are believed to
be due to the orbital motion of matter in an accretion disk,
 raised hopes concerning observational constraints on the equation of
 state (EOS) of matter at supranuclear densities (Kaaret et al. 1997,
 Klu{\'z}niak 1998,
Zhang et al. 1998,
 Miller et al. 1998, Thampan et al. 1999, Schaab \&
Weigel 1999).
 General relativity predicts the existence of the
marginally stable (MS) orbit, within which no stable circular
motion is possible. This implies the existence of the innermost
stable circular orbit (ISCO) around neutron stars.
 The frequency of the ISCO is
an upper bound on the frequency of stable orbital motion around
neutron stars. Whether the ISCO is separated from neutron star
surface by a gap, or its radius coincides with stellar equatorial
radius, depends on the star mass and on the EOS of neutron star
matter. On the other hand, accreting neutron stars in LMXBs are
expected to be rotating, and this influences both neutron star
structure and the ISCO. Therefore, in order to attempt to use
observed frequencies of QPOs to constrain the EOS of dense matter,
one has to calculate the ISCO as a function of stellar mass and
stellar rotation frequency. Such a procedure is based on the
assumption that the observed upper QPO frequency is due to orbital
motion, and that the effects of magnetic field, accretion, and
radiation drag on the matter flow can be neglected.

A basic assumption of the present paper is that the frequency of the
upper kHz QPO is the orbital frequency of the inner edge of an
accretion disk  surrounding the compact object, which will be identified
with the ISCO. This is the leading interpretation of the QPOs. However,
alternative models of the kHz QPOs were also proposed. In a model of
Alpar \& Yilmaz (1997) the kHz QPOs are explained in terms of
wave packets of sound waves  in the inner disk. In a series of papers,
Titarchuk and collaborators propose a model in which the QPOs result
from radial oscillations of the plasma in the boundary layer, i.e.
in the region between the ISCO and stellar surface
(see Titarchuk \& Osherovich 1999 and references therein).
These alternative
models will not be considered in our study.

\begin{figure*}     
 \resizebox{\hsize}{!}{\includegraphics[angle=-90]{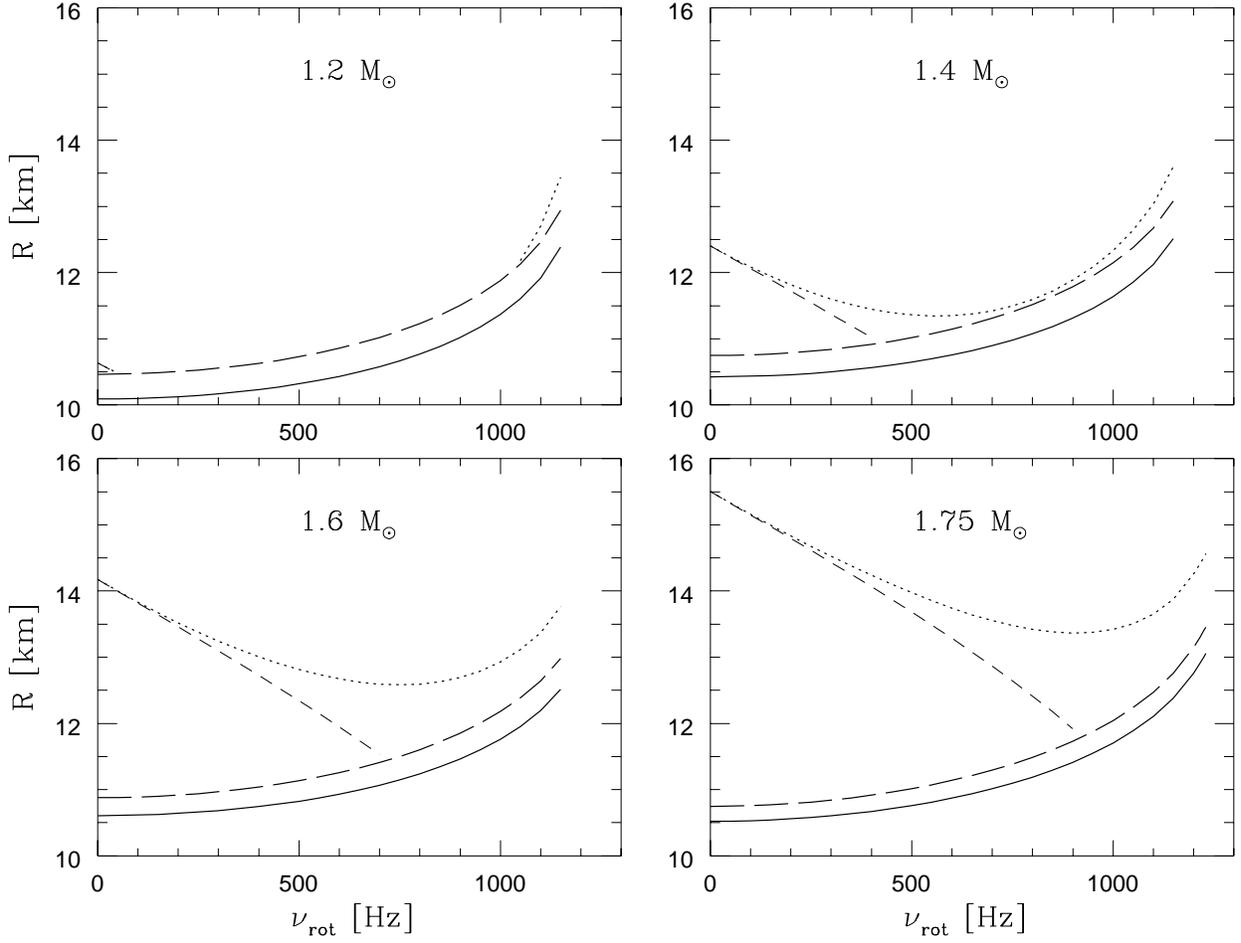}}
\caption{ The radius of the ISCO (dotted line), equatorial radius
of a bare strange star (solid line), equatorial radius of a
strange star with maximally thick solid crust (long-dashed line),
versus rotation frequency of strange star. Thin short-dashed  line
corresponds to slow-rotation approximation of Klu{\'z}niak \&
Wagoner (1985). Calculations  were performed for the SQM1 EOS of
strange matter (see the text). Figures  correspond to stellar
models with  fixed total baryon number,
 equal to that of a static star of  gravitational mass
of $1.2,~1.4,~1.6, 1.75~{\rm M}_\odot$. Maximum mass for static
strange stars is $1.8~{\rm M}_\odot$ } \label{fig1}
\end{figure*}

In the present paper we describe  results of exact  calculations of
the ISCO, under the assumption that the compact object is not a
neutron star, but a strange star. A strange star is
 composed of self-bound quark matter,
 which at zero pressure would constitute a real ground state of matter (strange
 matter), with energy per unit
baryon number lower  than that of $^{56}{\rm Fe}$ crystal (Witten
1984, Farhi \& Jaffe 1984, Haensel et al. 1986, Alcock et al.
1986; for a recent review of physics and astrophysics of strange
matter, see Madsen 1999). Recently, strange stars were invoked by several
authors in the context of modeling of observational properties of
some X-ray and gamma-ray sources (Bombaci 1997, Cheng et al. 1998,
Dai \& Lu  1998, Li et al. 1999). First  study of possibility
of existence of strange stars  in LMXBs  exhibiting  kHz QPOs was restricted
to slow-rotation approximation for the ISCO, neglected the effect of
rotation on the strange star  structure, and used simplified EOS of strange
matter, with massless, non-interacting quarks (Bulik et al. 1999
). Very recently, the ISCOs around bare strange stars were calculated,
assuming a simplified EOS of strange matter,  for
the limiting case of rotation at Keplerian frequency (Stergioulas
et al. 1999). In both these studies the possible presence of the solid crust on
the strange star  surface was not taken into account.

 In principle, a strange star  could be covered
 by a thin   crust of normal matter, a possibility which is
particularly natural in the case of LMXB. The problem of formation
and structure of a crust on an accreting  strange star  was
studied by  Haensel \& Zdunik (1991)(see also Miralda-Escud{\'e}
et al. 1990). Because of its low mass, typically $\la 10^{-5}~{\rm
M}_\odot$, the effect of the crust on the exterior spacetime is
negligible. However,  it determines the location of the star
surface, due to its finite thickness of $\sim 200-300~$m. The
matter distribution within the strange core, relevant for the
exterior metric of rotating strange star, is characterized by a
very flat density profile: for a massive strange star, density at
the stellar center is typically only 2-3 times larger than that at
the outer edge of the strange core. This has to be contrasted with
the density distribution within a massive neutron star, which
decreases continuously from $\sim 10^{15}~{\rm g~cm^{-3}}$ at the
center to a few ${\rm g~cm^{-3}}$ at the surface.
 The differences between the density profiles of a strange star
and a neutron star result from the
basic difference in the EOS of their interiors.
In the case of a rapidly rotating compact object (situation relevant to
 LMXB), the differences between matter distributions within neutron star
 and strange star
may be  expected to  imply differences in the spacetime exterior
to the compact object, and in particular, differences in
the properties of the ISCO.
\section{ISCOs around strange stars for
standard MIT Bag Model of strange matter}
Our EOS of strange matter, composed of massless u, d quarks, and
massive s quarks, is based on the MIT Bag Model. It involves three
basic parameters: the bag constant, $B$, the mass of the strange
quarks, $m_{\rm s}$, and the QCD coupling constant, $\alpha_{\rm
c}$ (Farhi \& Jaffe 1984, Haensel et al. 1986, Alcock et al.
1986). Our basic  EOS corresponds to standard values of the Bag
Model parameters for strange matter:  $B=56~{\rm MeV/fm^3}$,
$m_{\rm s}=200~{\rm MeV/c^2}$, and $\alpha_{\rm c}=0.2$ (Farhi \&
Jaffe 1984, Haensel et al. 1986, Alcock et al. 1986).  This EOS of
strange quark matter will be hereafter referred to as SQM1. It
yields energy per unit baryon number at zero pressure $E_0=918.8
~{\rm MeV} <E(^{56}{\rm Fe})=930.4~$MeV. For the SQM1 EOS maximum
allowable mass for static strange star  models is $M_{\rm
max}^{\rm stat}=1.8~{\rm M_\odot}$.

The general relativistic models of stationary rotating strange stars
have been calculated by means of the multi-domain spectral method,
developed recently by Bonazzola et al. (1998). Details of the
calculation method, specifically adapted for rotating strange stars, may be
found in Gourgoulhon et al. (1999). Having calculated a particular
stationary rotating strange star model, and its exterior
spacetime, we  determine the frequency of a particle in stable
circular orbit in the equatorial plane, $\nu_{\rm
orb}(r)$, where $r$ is the radial  coordinate of the orbit. By
testing the stability of orbital motion, we determine the radius of
the innermost, marginally stable orbit, $R_{\rm ms}$, and its
frequency $\nu_{\rm ms}$ (see, e.g., Datta et al. 1998, for
the equations to be solved).  Our numerical code calculating the ISCO
has been successfully tested  by comparing our results
for the polytropic $\gamma=2$ EOS with those obtained by
Cook et al. (1994a). Let us notice that the high precision of our
numerical method  makes it particularly suitable for the
determination of $R_{\rm ms}$, which requires calculation of
second derivatives of metric functions: these latter are better
evaluated by the spectral method we employ than by means of finite
differences.

 No orbital motion is possible for $r<R_{\rm ms}$.
The values of $R_{\rm ms}$ and $\nu_{\rm ms}$ for particles
corotating with strange star  differ from those for counterrotating ones.
In the present paper we restricted ourselves to the corotating
case, relevant for the LMXB. We neglect the effect of magnetic field,
accretion, and radiation drag  on the location of the ISCO, which
is justified for $B\la 10^{8}~$G and ${\dot M}\ll {\dot M}_{\rm
Edd}$.

Let us consider a strange star, rotating at a frequency $\nu_{\rm
rot}$, with equatorial radius $R_{\rm eq}$. If $R_{\rm ms}>R_{\rm
eq}$, then stable orbits exist for $r>R_{\rm ms}$; the ISCO has
then the radius $R_{\rm ms}$ and the frequency $\nu_{\rm ms}$, and
there is a gap of width $R_{\rm ms}-R_{\rm eq}$ between the ISCO
and the strange star  surface. However, if $R_{\rm ms}<R_{\rm
eq}$, then $R_{\rm ISCO}=R_{\rm eq}$, $\nu_{\rm ISCO}= \nu_{\rm
orb}(R_{\rm eq})$; and the accretion disk extends then down to the
strange star surface (or, more precisely, joins stellar surface
via a boundary layer).

While the exterior spacetime
 is  practically not influenced by the presence
of a solid crust on the strange star surface, the value of $R_{\rm eq}$ is
affected by it.
Neutrons are
absorbed by strange matter, and therefore the density at the
bottom of the crust, $\rho_{\rm bott.cr.}$,
 cannot be higher than $\rho_{\rm n-drip}\simeq
4\times 10^{11}~{\rm g~cm^{-3}}$ (lower values of $\rho_{\rm
bott.cr.}$ were discussed by Huang \& Lu 1997). The equatorial
thickness of the crust, $t_{\rm eq}$, which we calculate,
corresponds to $\rho_{\rm bott.cr.} =\rho_{\rm n-drip}$, and is
therefore an upper bound on $t_{\rm eq}$. At a fixed baryon
number, rotation increases $t_{\rm eq}$, as compared to the value
for a static strange star, $t_0$. Dependence of $t_{\rm eq}$ on
$\nu_{\rm rot}$ is well described by a formula $t_{\rm
eq}(\nu_{\rm rot})=t_0\cdot [1+0.7 (\nu_{\rm rot}/\nu_{\rm
K})^2]$, where $\nu_{\rm K}$ is the Keplerian (mass shedding)
frequency of strange star. For rotating strange stars  our formula
for $t_{\rm eq}$ reproduces numerical results of Glendenning \&
Weber (1992) within better than 2\% in all cases considered by
these authors.
 It is obvious that the bare strange star rotating
at Keplerian limit would be unstable if we added a crust of 
nonzero thickness due to the increase of the radius,  leaving the
mass and the angular momentum practically unaltered. Thus at a
fixed baryon mass of rotating strange star, the presence of the
crust implies a decrease of the Keplerian frequency. Knowing the
dependence of the radius of the strange core and rotational
frequency on the stellar  angular momentum one can estimate the
point of the Keplerian instability for a strange star with crust.
%
\begin{figure}     
 \resizebox{\hsize}{!}{\includegraphics[angle=-90]{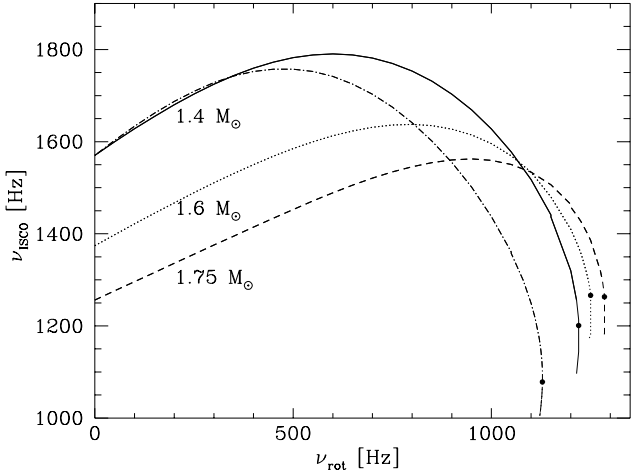}}
\caption{Frequency of the ISCO versus the rotation frequency of strange
star,  for the same SQM1 EOS of strange matter.
Dash-dotted line was obtained for a simplified
 SQM0 model of strange matter with massless, noninteracting quarks
  (see the text). Each curve corresponds to a fixed
baryon mass, equal to that of a static strange star of
gravitational mass indicated by a label.
Along each curve, angular momentum increases from $J=0$ (static
configuration) to $J_{\rm max}$ (Keplerian limit). Filled circles
 correspond to Keplerian configurations of strange stars with
 crust. Segments below the filled circles can be reached
only by the bare
 strange stars.}
 \label{fig2}
\end{figure}

Let us consider  first rotating strange stars  for the SQM1 EOS, which
corresponds to the ``standard set'' of the Bag Model parameters for
strange matter. A sample of our results for sequences of
rotating strange star models with fixed baryon number are presented in Fig. 1.

  The form of Fig. 1 is analogous to that constructed by Miller et al.
(1998) for neutron stars, and therefore is suitable for discussion
of the differences between neutron stars and strange stars. For
strange stars  with static mass $M\ga 1.4~{\rm M}_\odot$, we have
always $R_{\rm ms}>R_{\rm eq}$, for any $\nu_{\rm rot}$. So, for
$M\ga 1.4~{\rm M}_\odot$ the gap between strange star  surface
(with or without solid crust) and the ISCO exists at any strange
star  rotation rate. Even for lower $M$, the gap, which disappears
at moderate rotation rates, reappears at  $\sim$ 1 kHz frequency
of rotation; this is visualized by the $M=1.2~{\rm M}_\odot$ case
in Fig. 1. At $\nu_{\rm rot}= \nu_{\rm K}$, the ISCO is always
separated from the strange star surface  by a gap. Clearly, these
features of the ISCO around strange stars are quite different from
those obtained  by Miller et al. (1998) for neutron stars  with
the FPS EOS (see their Fig. 1). Note that the existence of a gap
($R_{\rm ms}>R_{\rm eq}$) is expected to lead to a qualitatively
different spectrum of X-ray radiation from LMXB, compared to the
no-gap case (Klu{\'z}niak et al. 1990).

\begin{figure*}     
 \resizebox{\hsize}{!}{\includegraphics[angle=-90]{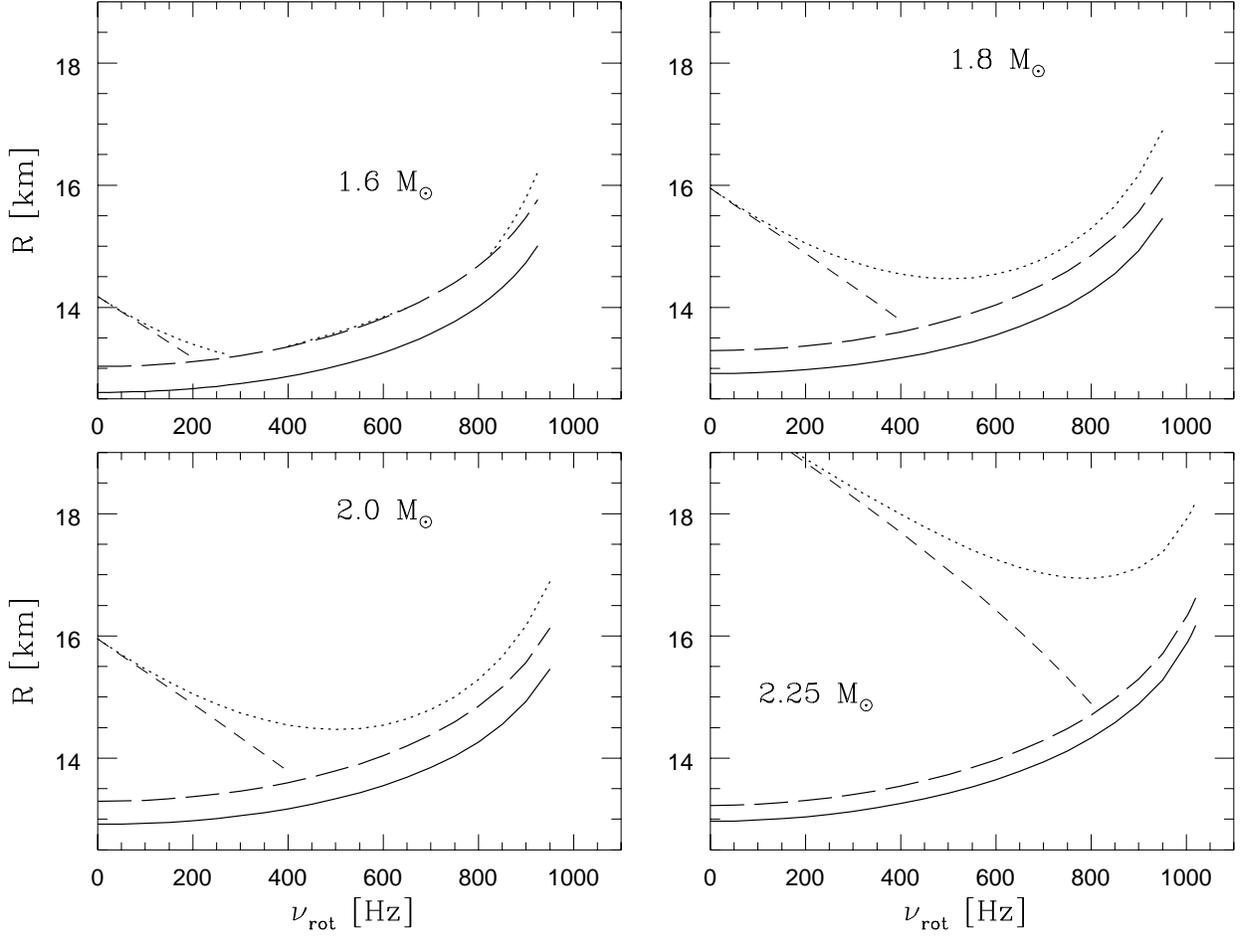}}
\caption{ The radius of the ISCO (dotted line), equatorial radius
of  bare strange star (solid line), equatorial radius of
strange with  a solid crust (long-dashed line),
versus rotation frequency of strange star, for the SQM2 EOS of
strange quark matter (see the text). Thin short-dashed line
corresponds to slow-rotation approximation of Klu{\'z}niak \&
Wagoner (1985).  Figures correspond to stellar models with  fixed
total baryon number,
 equal to that of a static star of the gravitational mass
of $1.6,~1.8,~2.0,~2.25~{\rm M}_\odot$. Maximum mass of static
strange stars is $2.3~{\rm M_\odot}$} \label{fig3}
\end{figure*}

Constraints on EOS of dense matter, resulting from  the
kHz QPOs observations, were initially derived within the slow-rotation
approximation,
in which $R_{\rm ms}\simeq R_{\rm ms}^{\rm s.r.}=6GM/c^2\cdot[1-(2/3)^{3/2}j]$,
 $j\equiv Jc/GM^2$, and $J$ is stellar angular momentum (Klu{\'z}niak
\& Wagoner 1985). However as pointed out by Shibata \& Sasaki
(1998), the mass quadrupole moment is as important as the angular
momentum in determining the ISCO. Indeed as we can see in Fig. 1,
slow rotation approximation yields $R_{\rm ms}^{\rm s.r.}$, which
in the case of $M=1.4~{\rm M}_\odot$ diverges from exact $R_{\rm
ms}$ for $\nu_{\rm rot} \ga 500~$Hz. Moreover, for rotating
strange stars
 $R_{\rm ms}^{\rm s.r.}$ leads always to
disappearance of the gap at sufficiently high $\nu_{\rm rot}$,
 in contrast  to exact calculation.

A quantity of particular interest in the context of the interpretation of
observed kHz QPOs in LMXB, is the
maximum frequency of the stable circular orbit  at a given $\nu_{\rm rot}$,
 which we identify here with that of the ISCO. In principle, both
$\nu_{\rm ISCO}$ and $\nu_{\rm rot}$ are observable (measurable)
quantities, which can be thus used for confronting stellar models
with observations.  In Fig. 2 we present curves $\nu_{\rm
ISCO}(\nu_{\rm rot})$ for the SQM1 EOS. As in Fig. 1, baryon
masses are fixed along each curve, while labels correspond to
gravitational mass of non-rotating strange star.

In all cases, displayed in Fig. 2, gap between stellar surface and
ISCO exists,  and therefore $\nu_{\rm ISCO}=\nu_{\rm ms}$. The
dash-dotted  line was calculated for a simplified EOS, with
massless, non-interacting quarks ($m_{\rm s}=0$, $\alpha_{\rm
c}=0$, $B=56~{\rm MeV/fm^3}$), hereafter referred to as SQM0 (such
a type of the EOS of strange quark matter was used in Bulik et al.
1999). For $\nu_{\rm rot} \ga 500~$Hz, neglecting strange quark
mass (and, to a smaller extent, neglecting QCD interactions) leads
to a rather severe underestimate of $\nu_{\rm ISCO}$ for rapidly
rotating strange stars (by 200 Hz at $\nu_{\rm rot}=1~$kHz). As we
stressed before, presence of the solid crust does not influence
the space-time outside rotating strange strange star. However,
solid crust decreases (by about 10\%) the value of $J_{\rm max}$
of strange stars of a given baryon mass. Complete curves in Fig. 2
correspond to bare strange stars.  Rotating configurations with
crust terminate at filled dots, corresponding to the Keplerian
limit in the presence of the crust. The effect of the presence of
the crust on the value of $\nu_{\rm ISCO}$ at $\nu_{\rm K}$ turns
out to be significant, which is due to the steepness of the
$\nu_{\rm ISCO}(\nu_{\rm rot})$ curve for bare strange stars at
$J\simeq J_{\rm max}$. For $J$ approaching $J_{\rm max}$ bare
strange star undergoes strong deformation with increasing $J$.
This deformation in turn implies strong decrease of $\nu_{\rm
ISCO}$ with increasing rotation frequency. Consequently, the
values of $\nu_{\rm ISCO}$ for maximally rotating strange stars
with crust is about hundred Hz higher than for bare strange stars.
This effect increases
 with decreasing strange star mass. At fixed $B$, the
effect is stronger for the EOS which produces less compact strange
stars of a given mass. Therefore, it  is strongest for the SQM0
EOS with massless, noninteracting quarks, where maximally rotating
configurations of $1.4~{\rm M}_\odot$ with crust have the ISCO
frequency of 1.1 kHz, to be compared  with less than 1 kHz for
maximally rotating bare strange stars.

The problem of  an appropriate parametrization of the one-parameter family
of rotating strange stars with fixed  baryon mass deserves a comment.
These configurations may be labeled by the value of the total
angular momentum $J$, which changes from $J=0$ in the static case to
 $J_{\rm max}$ at the Keplerian limit. As one can see in Fig.2,
for bare strange stars, Keplerian configuration is not that with
maximum $\nu_{\rm rot}$. The reason is that for very rapidly
rotating strange stars the increase of the total angular momentum
results mainly in the oblateness of the configurations leading to
the significant increase of the equatorial radius without an
increase of $\nurot$ (or even with a decrease of $\nurot$ very
close to $J_{\rm max}$). As a consequence at fixed baryon mass the
Keplerian configurations is reached not due to the increase of
$\nurot$ but because of the increase of the equatorial radius
related to the deformation of the star. It is worth noticing that
the difference between $\nu_{\rm rot,max}$ and $\nu_{\rm K}$ is of
the order of one percent. The existence of this difference
 implies that for $J\simeq J_{\rm max}$
 it is in principle possible to spin up
the strange star by the angular momentum loss. Such a  situation was
previously discussed in the case of supramassive neutron stars
(Cook et al. 1994b) and supramassive
strange stars (Gourgoulhon et al. 1999).
\section{Confronting the standard MIT Bag Model of strange matter with
QPO observations}

Let us pass now to the confrontation of our results for strange
stars with observations of the QPOs. Nearly twenty LMXBs,
exhibiting QPOs, have been observed (van der Klis 2000). The
upper-peak frequency, $\nu^{\rm u.p.}_{\rm QPO}$, is usually
interpreted as the frequency of the orbital motion around a
neutron star. The most general observational constraint on a
neutron star  in LMXB is thus $\nu^{\rm u.p.}_{\rm QPO} \le
\nu_{\rm ISCO}$. Highest observed $\nu^{\rm u.p.}_{\rm
QPO}$ is {\chng $1329\pm 4$ Hz in 4U 0614+09 (van Straaten et al. \cite{Straaten}). 
Condition $\nu_{\rm ISCO}\ge 1.33$~kHz is satisfied by nearly all strange star
models displayed in Fig. 2 (except those rotating very close to 
the Keplerian frequency and slowly rotating maximum mass model). 
In particular for the spin frequency of the star $\nu_{\rm spin}=312$~Hz
(Ford et al. \cite{Ford97}, van Straaten et al. \cite{Straaten}) all
stellar configurations for SQM1 model of strange matter are allowed.} 

For neutron stars, condition $\nu_{\rm ISCO}\ge 1.2$~kHz 
{\chng (considered by Thampan et al. 1999 as the highest 
$\nu^{\rm u.p.}_{\rm QPO}$)}
eliminates stellar masses below some limit, ranging from 
$0.6~{\rm M_\odot}$ for
stiff EOS to $1.4~{\rm M_\odot}$ for soft EOS (Thampan et al.
1999). In this case the innermost allowed orbit is defined by the
radius of the star and corresponds to the Keplerian frequency
at the surface $\nu_{\rm K}$. {\chng This conclusion would be stronger 
in the case of $\nu^{\rm u.p.}_{\rm QPO}=1.33$~kHz excluding the
softest EOS and shifting the above mass limits to a little higher values
(see Fig 1. in the paper by Thampan et al. 1999).} 
Such a constraint does not apply to bare strange stars, for which in
the limit of $M\ll {\rm M_\odot}$ one gets $\nu_{\rm K}\simeq
(G\rho_{\rm sm}/3\pi)^{1/2}= 0.841\cdot (\rho_{\rm
sm,14})^{1/2}$~kHz, where $\rho_{\rm sm,14}$ is the density of
strange matter at zero pressure, in the units of $10^{14}~{\rm
g/cm^3}$. For reasonably high values of $\rho_{\rm sm}$, in
particular for those considered in the present paper, one gets
{\chng $\nu_{\rm K}>1.33$~kHz } for low-mass, 
slowly rotating bare strange stars.
In the case of strange stars with  crust, condition $\nu_{\rm
K}>1.33$~kHz turns out to be  violated  for $M\la 0.4~{\rm M}_\odot$.

The behavior of QPOs in 4U 1820-30 has been interpreted as
evidence for $\nu^{\rm u.p.}_{\rm QPO}=\nu_{\rm ms}$ in this LMXB
(Kaaret et al. 1999, and references therein). Accepting such an
interpretation of $\nu^{\rm u.p}_{\rm QPO}= 1.07$ kHz implies
strong constraints on neutron star  model, and therefore, on the
neutron star EOS
(Klu{\'z}niak 1998, Miller et al. 1998). Only a few existing EOS
of neutron star  matter allow simultaneously for $R_{\rm eq}<R_{\rm ms}$ and
$\nu_{\rm ms}=1.07~$kHz. It is clear  from Fig. 2 that
SQM1 models of strange stars cannot give $\nu_{\rm ISCO}$ as
low as 1.07 kHz at slow rotation rates. In the case of the SQM0
EOS one is able to get such low $\nu_{\rm  ISCO}$ for bare strange
stars of $M\la  1.4~{\rm M}_\odot$ rotating close to Keplerian
frequency; we confirm in this way result of Stergioulas et al.
(1999). However, as we see in Fig. 2, passing to an EOS which at
the same value of $B$  includes effects of strange quark mass and
of lowest order QCD interaction increases the values of $\nu_{\rm
ISCO}$ of bare strange stars at high rotation rates to such
extent, that the value of $1.07$~kHz cannot be
reproduced.The presence of solid crust on  rotating
strange stars described by the ``standard strange matter EOS''
SQM1 excludes $\nu_{\rm ISCO}(\nu_{\rm K})$ lower than 1.2 kHz. In
the case of the simplest SQM0 EOS the value of $\nu_{\rm
ISCO}(\nu_{\rm K})$ is increased by the presence of the crust a little
above 1.07 kHz. Generally, the presence of the crust on rotating
strange star with standard strange matter EOS, such as SQM1 (or
SQM0) excludes possibility of getting $\nu_{\rm ISCO}$ as low as
1.07 kHz for any  possible  rotation rates.
\section{MIT Bag Model of strange matter consistent with
QPO observations}

In order to get $\nu_{\rm ISCO}$ as low as 1.07 kHz at slow and moderate
rotation rates, one has
to consider a specific  set of the MIT Bag Model parameters,
 characterized by significantly lower values of both $B$ and
$m_{\rm s}$, and higher value of $\alpha_{\rm c}$, than those
characteristic of the  SQM1 model. In this way one is  able to
increase significantly the value of $M_{\rm max}^{\rm stat}$, and
get ISCO frequencies as low as 1 kHz at slow rotation rates.
For such a choice of EOS it is also relatively easy to
get $\nu_{\rm ISCO}\la 1$~kHz for a broad range of masses of
configurations rotating close to the Keplerian limit
 (see below). An example of
such an  EOS, hereafter referred to as the SQM2, was obtained assuming
$B=40~{\rm MeV/fm^3}$, $m_{\rm s}=100~{\rm MeV/c^2}$, and
$\alpha_{\rm c}=0.6$. At zero pressure, the SQM2 model yields
energy per unit baryon number $E_0=874.2$~ MeV. Let us stress
that despite the relatively  low value of $B$, the standard condition
that  neutrons do not fuse (coagulate) spontaneously into
strangelets (droplets of quark matter),  is satisfied by  this
model. Maximum mass of static strange stars  for the SQM2 EOS is $2.3~{\rm
M}_\odot$.

\begin{figure}     
 \resizebox{\hsize}{!}{\includegraphics[angle=-90]{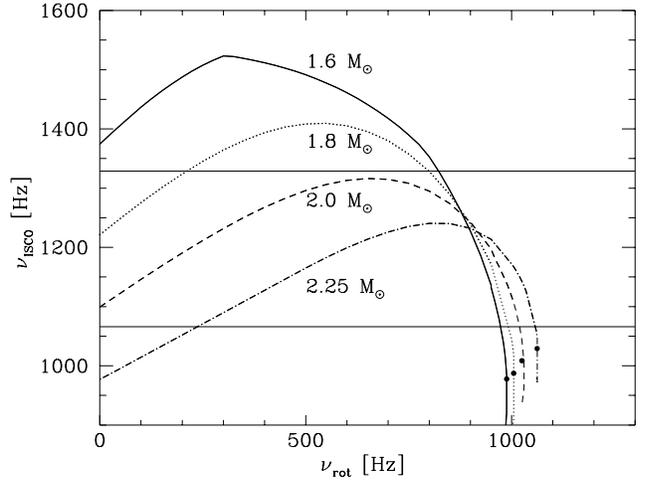}}
\caption{Frequency of ISCO versus rotation frequency of strange star
 for the SQM2 EOS  of strange matter (see the text).
Lines correspond  to stellar
models with  fixed total baryon number,
 equal to that of a static star of the gravitational mass
of $1.6,~1.8,~2.0, ~2.25~{\rm M}_\odot$,
 Lower thin horizontal line:
 upper-peak frequency observed in 4U 1820-30, 1.07~kHz. 
Upper horizontal line:
 maximum  upper-peak frequency  observed in the QPOs
in LMXBs, 1.33~kHz. Keplerian configurations for strange stars with crust
are indicated by filled circles. Segments of curves below filled circles
correspond to bare strange stars. Maximum mass for static strange stars
is 2.3~${\rm M}_\odot$.  } \label{fig4}
\end{figure}
%

Our SQM2 model is a rather extreme one, as far as the
values of the $B$, $m_{\rm s}$, and $\alpha_{\rm c}$ parameters
are concerned. Canonical value of $B$, resulting from fitting
hadronic masses, is $59~{\rm MeV/fm^3}$ (De Grand et al. 1975),
significantly higher than  $B=40~{\rm MeV/fm^3}$  used in the SQM2
model. On the other hand, $m_{\rm s} = 100~{\rm MeV/c^2}$ of the
SQM2 model is on the lower side of usually considered  $m_{\rm s}$
values (Farhi \& Jaffe 1984, Madsen 1999). Finally, $\alpha_{\rm
c}=0.6$ is on the upper side of the interval of the $\alpha_{\rm
c}$ values considered in the strange matter calculations
(Farhi \& Jaffe 1984).

Our results for the SQM2 EOS, analogous to those displayed in Fig.~1
and Fig. 2 for the  SQM1 EOS, are shown in Fig.~3 and Fig.~4.
The main differences between these models can be explained by the
scaling laws with the bag constant, discussed in
Sect. 5. The features of $R_{\rm ms}$ and radius
of the rotating strange star of given mass for SQM1 model corresponds to
the star SQM2 with the mass larger by the
factor $\sim (B_1/B_2)^{1/2}$.
 As one can see in Fig.~4, slowly rotating strange stars
 can have ISCO frequencies as low as $1-1.1$ kHz,
provided their mass is sufficiently high, $M\simeq 2.2-2.3~\msol$
just because maximum allowable mass
for static strange stars is sufficiently high. Moreover,
for bare strange stars,
the ISCO frequency below  1~kHz can also be reached for
 very rapid rotation close to the Keplerian limit. Less massive is
bare strange star, lower is the ISCO frequency reached at the
Keplerian limit. The presence of the solid crust makes
the window (subset)  of rapidly rotating configurations
allowing for
$\nu_{\rm ISCO}=1.07$~kHz significantly narrower.
 These configurations are very close to the Keplerian ones.
Notice that the SQM2 EOS is simultaneously consistent with
$\nu_{\rm ISCO}\ge 1.33$~kHz, provided the strange star mass $M\la 1.8~{\rm
M}_\odot$.
\section{Discussion and conclusion}
The features of ISCOs around rapidly rotating
strange stars, described
in the present paper
 for a particular choice of strange matter EOS, are actually
generic.
The MIT Bag Model EOS of strange quark matter depends on $B$, $m_{\rm s}$,
and $\alpha_{\rm c}$ in a way, which implies specific scaling properties
with respect to change of $B$ (Haensel et al. 1986, Zdunik \& Haensel 1990).
 As a consequence, the global parameters of rotating strange stars  scale with
some power of $B$, which allows one to determine the values of $M$,
$R_{\rm eq}$, $R_{\rm ms}$, etc., for $B$, $\alpha_{\rm c}$, and
$m_{\rm s}$, from those calculated
 for $B_0$, $\alpha_{\rm c}$ and strange quark mass
 $m_{\rm s}(B_0/B)^{1/4}$. All  length-type quantities
(stellar radius, thickness of the
crust and radius of the ISCO)  scale as
$B^{-1/2}$, and all frequencies ($\nu_{\rm ISCO}$, $\nu_{\rm rot}$)
scale as
 $B^{1/2}$, e.g.,
$\nu_{\rm ISCO}[B]=\nu_{\rm ISCO}[B_0]\cdot (B/B_0)^{1/2}$ and
$R_{\rm ms}[B]=R_{\rm ms}[B_0]\cdot (B/B_0)^{-1/2}$. Thus, for
other values of $B$ the patterns of lines in Figs. 1-4 do not
change, provided one rescales the axes and stellar masses.

Our calculations show that the properties of the ISCOs around strange stars
differ from those around neutron stars.
A generic property is the existence of the
gap between the ISCO and the stellar surface, for both slowly and rapidly
rotating strange stars.

The highest observed QPO frequency of {\chng 1.33 kHz in 4U
0614+91} can be easily interpreted as an orbital frequency around
strange star based on the standard SQM1 EOS of strange matter,
with no significant constraint on strange star mass and rotation
rate. In the case of the SQM2 EOS, the orbital origin of the {\chng 1.33
kHz QPO implies $M\la 1.8~{\rm M}_\odot$ at rotation frequencies
$\sim 300$~Hz}, while frequencies close to the mass shedding limit are
excluded.  

The value of the ISCO frequency at the
Keplerian limit is significantly influenced by the presence of a
crust  on the strange star
surface, which increases this frequency
 by about  hundred Hz compared to the value for
a bare strange star of the same mass.
As one expects the presence of
a crust on a strange star in a LMXB, we conclude that only
slowly rotating strange
stars with  mass above $2.2~{\rm M}_\odot$ seem to be consistent
with $\nu_{\rm ISCO}=1.07$~kHz. This excludes EOS of strange matter
corresponding to the standard bag model parameters, and can be
satisfied only by choosing a set of parameters quite different
from the standard one.

The numerical results discussed in the present paper show that
consistency of the ISCOs around slowly rotating strange stars
 with orbital-motion
interpretation of QPOs in LMXBs can be achieved only with a
substantial tuning of the MIT Bag Model parameters of strange
matter.
Our SQM2 EOS is a result of such a tuning. For this EOS,
the condition $\nu_{\rm ISCO}\simeq 1$~kHz is satisfied not only
for slowly rotating massive models with $M\ga 2.2~{\rm M}_\odot$,
but also for a broad range of masses of configurations close to
the Keplerian  limit.
In contrast to $\nu_{\rm ISCO}$ at low rotation rate,
which decreases with increasing baryon mass, the ISCO frequency at the
Keplerian limit decreases with decreasing baryon mass of rotating
strange star.

\begin{acknowledgements}
During his stay at DARC, Observatoire de Paris, P. Haensel was
supported by the PAST professorship of French MENRT. This research
was partially supported by the KBN grants No. 2P03D.014.13, 2P03D.021.17.
The numerical calculations have been performed on computers purchased
thanks to a special grant from the SPM and SDU departments of
CNRS. We are very grateful to the referee, P. Kaaret, for helpful
comments and suggestions, which influenced the final version of
the present paper.
\end{acknowledgements}

\end{document}